\documentclass[aps,pra,twocolumn,tightenlines,amsmath,amssymb,superscriptaddress, 10pt]{revtex4-2}

\usepackage{graphicx} 
\usepackage[utf8]{inputenc}
\usepackage{siunitx}
\usepackage{array}
\usepackage{dsfont}
\usepackage{booktabs}
\usepackage[english]{babel}			
\usepackage{xcolor}
\usepackage{amssymb,amsmath}		
\usepackage{url}					
\usepackage{marvosym}				
\usepackage[T1]{fontenc}            %
\usepackage{wrapfig}				
\usepackage{charter} 				
\usepackage{blindtext}				
\usepackage{datetime}				
\usepackage{lipsum}                 
\usepackage{float}                  
\usepackage{makecell, tabularx}
\usepackage{fancybox}  
\usepackage{dcolumn}
\usepackage{bm}
\usepackage{hyperref}

\definecolor{colorJ}{rgb}{.1,.8,.5}

\newcommand{\sparrow}{Sparrow Quantum Aps, Nordre Fasanvej 215, 2000 Frederiksberg, Copenhagen, Denmark}
\newcommand{\bochum}{Ruhr-Universitat Bochum, Universitatsstrasse 150, 44801 Bochum, Germany}

\begin{document}
\title{A photonic source with half-a-GHz single-photon flux}

\author{P.~Zahalka}
\affiliation{\sparrow}
\author{S.~Huijser}
\affiliation{\sparrow}
\author{A. Pancaldi}
\affiliation{\sparrow}
\author{S.~Kruger}
\affiliation{\bochum}
\author{X.~Zhao}
\affiliation{\sparrow}
\author{Z.~Liu}
\affiliation{\sparrow}
\author{I.~Suleiman}
\affiliation{\sparrow}
\author{R.~Jensen}
\affiliation{\sparrow}
\author{L.~Stefan}
\affiliation{\sparrow}
\author{A.~Ludwig}
\affiliation{\bochum}
\author{V.~Remesh}
\affiliation{\sparrow}
\author{J.~C.~Loredo}
\affiliation{\sparrow}
\author{P.~Lodahl}%
\affiliation{\sparrow}

\begin{abstract}
Advanced optical quantum technologies demands high quality quantum light generation at very high rates. Here, we report on a deterministic single-photon source  that simultaneously combines high excitation rates with high system efficiency to reach over {500 MHz} of in-fibre single-photon flux. The source delivers optical power of over {100~pW}, as is measured with an off-the-shelf powermeter, enabling a simple and direct way of determining the single-photon source fiber efficiency.
\end{abstract}

\maketitle

\section{Introduction}

Deterministic single-photon sources are essential components for high-efficiency photonic quantum information processing, quantum key distribution, optical quantum computing, and quantum enhanced machine learning \cite{UppuDetPhoton21,Chan2026}. The throughput of nearly every quantum photonic protocol scales linearly with the source clock-rate and decreases exponentially with the photon source efficiency. {Similarly, for} secure communication based on quantum key distribution, information sent through optical fibers is subject to losses and decoherence, which can be mitigated by boosting the repetition rate, avoiding the need for prohibitively long integration times. {These} applications fundamentally depend on a supply of pure and indistinguishable photons delivered at {rates that are directly} useful to the end-user---hence the demand for photonic sources with the highest single-photon flux. 



Single-photon sources based on solid-state quantum dot emitters can reach the highest photon fluxes by driving efficient sources at high repetition rates ~\cite{UppuScalable20,Tomm2021,maringVersatileSinglephotonbasedQuantum2024,dingHighefficiencySinglephotonSource2025,Loredo2026}. In this work, we drive a quantum dot single-photon source at excitation rates up to {1~GHz} to produce a fibred single-photon stream with over {500~MHz} photon flux, the highest value reported to date. The quantum light stream contains an optical power over $100$~pW, as measured directly with a standard optical powermeter.

\begin{figure}[htb!]
	\centering
	\includegraphics[width=.9\linewidth]{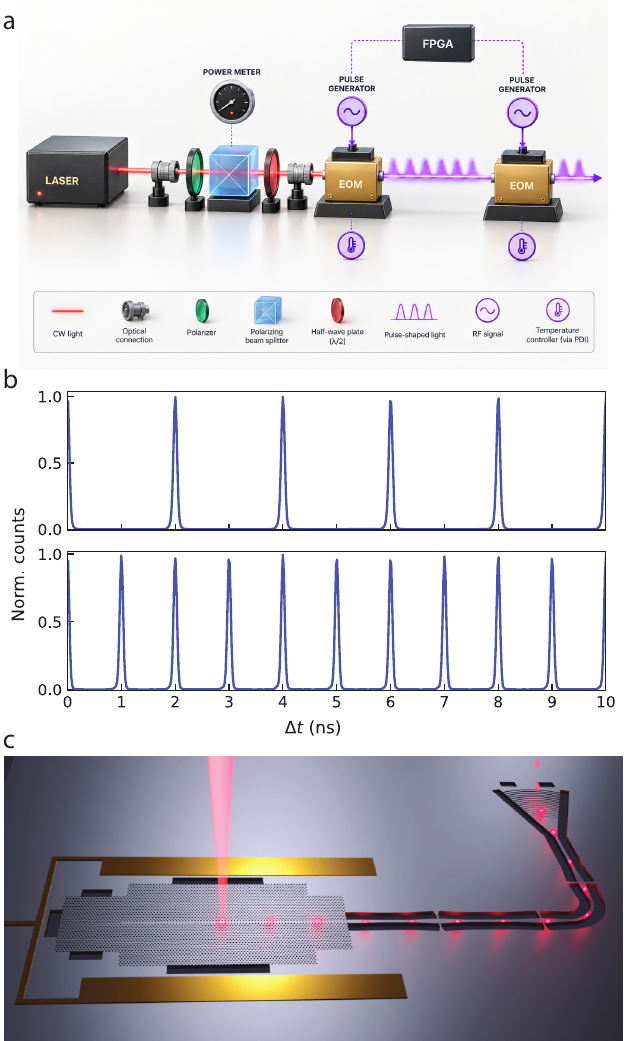}
    \caption{{\bf Pulse-carved excitation.} a) Two synchronised intensity EOMs transform an input CW laser to coherent laser pulses (pulse carving) with tunable repetition rate controlled by an FPGA (AI-generated concept image). b) Measured time traces of carved pulses at $500$~MHz (top) and $1$~GHz (bottom). c) High repetition rate carved pulses excite a quantum dot embedded in a PCW mode. The rate of single-photon production is maximum and limited by the inherent emitter decay-time.}
 \label{fig1}
\end{figure}

\begin{figure*}[htb!]
	\centering
	\includegraphics[width=1.\linewidth]{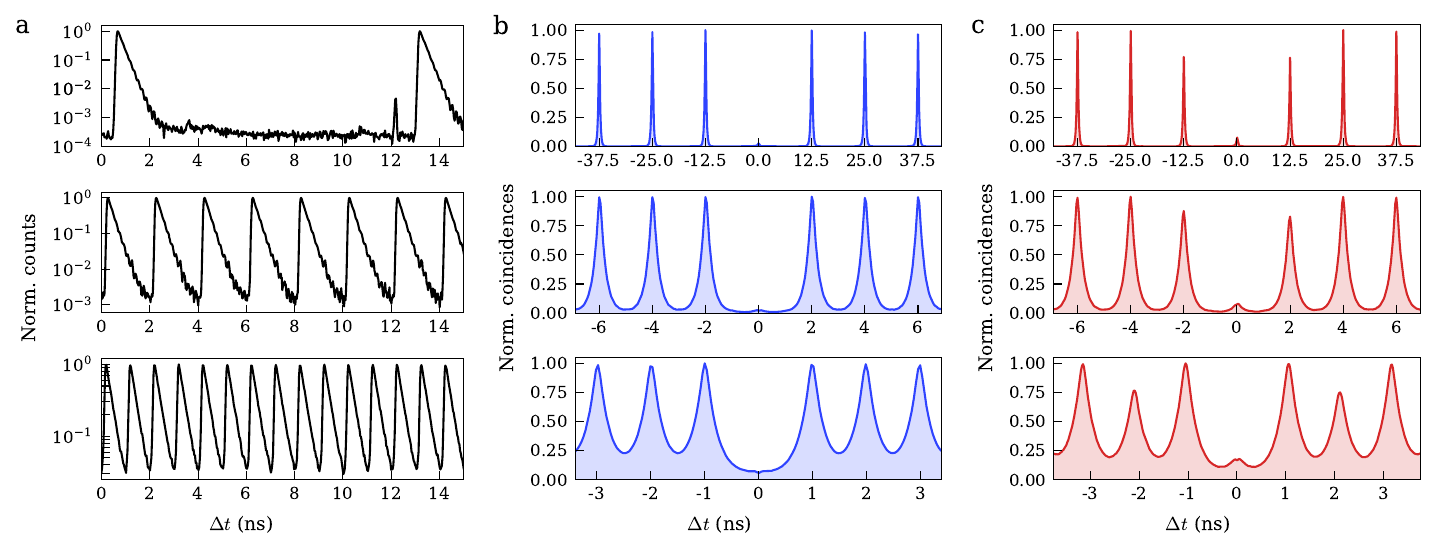}
    \caption{{\bf Single-photon metrics.} Single-photon measurements at $80$~MHz (top), $500$~MHz (middle), and $1$~GHz (bottom) excitation rates. a) Lifetime time traces, revealing a mono-exponential decay rate of $220$~ps. b) Second-order autocorrelation $g^{(2)}(\Delta t)$, from where we obtain $g^{(2)}(0)$ values of $(3.98{\pm}0.01)\%$, $(3.63{\pm}0.02)\%$, and {$(11.13{\pm}0.03)\%$}, respectively. c) Two-photon interference display uncorrected HOM visibilities $V_\text{HOM}$ of $(81.83{\pm}0.03)\%$, $(80.87{\pm}0.03)\%$, and $(55.14{\pm}0.03)\%$ . These values are obtained with no spectral filter in collection, and considering integration time windows of $2.2$~ns, $0.9$~ns, and $0.5$~ns. The decreasing values of purity and HOM visibility originate from increasing peak overlap at increasing excitation rates, which can be avoided by using sources with shorter lifetimes. When spectral filters are added in the collection path we obtain values of $g^{(2)}(0)=(2.80{\pm}0.03)\%$ and $V_\text{HOM}=(92.30{\pm}0.04)\%$ ($80$~MHz excitation), yet at the expense of $\sim12\%$ loss due to the optical transmission of the filters.}
 \label{fig2}
\end{figure*}

\section{Carved laser pulses}
We start by producing the necessary laser pulses at tunable repetition rates, for which we carve out pulses from a continuous-wave (CW) laser traversing high-bandwidth amplitude electro-optic modulators (EOMs)~\cite{Matthiesen2013,Dada2016,Meyer2023,Poortvliet2025}. {Figure~\ref{fig1}a} presents a sketch of the pulse carving setup. There, a wavelength tunable CW laser
with a linewidth below $1$~{kHz} is directed towards two cascaded temperature-controlled lithium-niobate intensity modulators. 
The EOMs are biased at their minimum-transmission point to maximize the extinction ratio and are driven by a time-varying voltage from a fast pulse generator, 
which in turn is triggered by a field programmable gate array (FPGA). The FPGA provides control over the pulse repetition rate and relative pulse timing, while the pulse generator determines the electrical pulse duration, adjustable between $40${~ps} and $100$~{ps}. The temperature of the EOMs is stabilized within $\SI{0.02}{\degree~C}$, ensuring a stable output laser power. The temporal profile of the resulting optical pulse train is characterized by a home-built linear autocorrelator, where we measure a pulse duration of ${40}$~ps and over ${40}$~dB extinction contrast at both ${80}$~MHz and ${1}${GHz} driving clock rates. This setup thus converts a time-static laser input to ${40}$~ps FWHM laser pulses triggered at a user defined rate, here tuneable up to $1$~GHz frequency{, see Fig.~\ref{fig1}b}. Practical advantages of this technique include wavelength and excitation clock tuneability, together with preserved coherence between excitation pulses inherited from the coherent laser input.

\section{Single-photon source}

We use the generated carved pulses to resonantly excite an InAs/GaAs quantum dot coupled to a photonic crystal waveguide (PCW){, see Fig.~\ref{fig1}c}. The emitter contains a neutral exciton state with an emission wavelength of {$\lambda{=}{933.1}$~nm} and a decay lifetime of {$\tau{=}{220}$~ps}. We deliberately opt for not placing any spectral filter as part of our collection optics, hence reporting the limits in performance and quality of our photon sources when all emitted light is considered, including the phonon sideband emission.

Figure~\ref{fig2}a shows the resulting single-photon time traces for excitation rates $R$ of $80$~MHz, $500$~MHz, and $1$~GHz, evidencing that the source is driven at maximum litefime-limited rate---no further time-orthogonal single-photon pulse fits in between existing pulses. Note that for an excitation rate of $R=500$~MHz, a single-photon time trace decays three orders-of-magnitude before a subsequent single-photon pulse, while for $R=1$~GHz the background overlap increases up to a $\sim4\%$ level. This indicates that negligible overlap between subsequent single-photon pulses can be obtained for ratios higher than $T/\tau=10$, given the separation between excitation laser pulses $T$ and single-photon lifetime $\tau$.

The driving laser pulses have a temporal duration of $\tau_\text{L}\sim40$~ps FWHM, a significantly long value relative to the single-photon lifetime of $\tau=220$~ps. Typically, resonant excitation of quantum emitters at these conditions results in increased multi-photon emission~\cite{Dada2016}, and accordingly a reduced single-photon purity. However, PCW quantum dot systems feature the possibility of destructive interference between laser leakage into the PCW mode and the multi-photon emission field~\cite{Gonzales:2025}, allowing to obtain high levels of single-photon purities even for these relatively long excitation pulse durations, see Fig.~\ref{fig2}b. Moreover, we perform Hong-Ou-Mandel (HOM) experiments to benchmark the limits of two-photon indistinguishability when using all unfiltered photon emission, see Fig.~\ref{fig2}c.



\begin{table*}[htp!]
  \centering
  \begin{tabular}{l l l l l l} 
    \hline
    {\bf Ref} & {\bf Source specification} & {\bf Efficiency} & {\bf Excitation method} & {\bf Rep. rate mechanism} \\
    \hline
     
   \cite{Anderson2020} & InAs/ InP, \SI{1515}{nm} & {---} & Pulsed ABE & Laser pulses, \SI{228}{MHz} \\
   \hline
   \cite{Yang2024} & InAs/InGaAs/GaAs, \SI{1530}{nm} & 3.59 MHz / 76 MHz & P-shell & Laser pulses, \SI{228}{MHz} \\
    \hline
      \cite{Hopfmann2021} & GaAs/AlGaAs, \SI{780}{nm} & {---} & TPE & Laser pulses, \SI{1}{GHz} \\
    \hline
      \cite{Rickert2024} & InAs/GaAs QDs, \SI{940}{nm} & {---} & P-shell, RF & Laser pulses, \SI{1.28}{GHz} \\
    \hline
     \cite{Rickert2025} & InAs/GaAs QDs \SI{940}{nm} & 1.2 MHz / 80 MHz & Off-resonant & Laser pulses, \SI{2.5}{GHz} \\
    \hline
    \cite{Dada2016} & InAs/GaAs QDs \SI{940}{nm} & 0.45 MHz / 160 MHz & RF & EOM-carving, \SI{160}{MHz} \\
    \hline
    \cite{Poortvliet2025} & InAs/GaAs QDs \SI{935}{nm} & {---} & RF & EOM-carving, \SI{80}{MHz} \\
    \hline
      \cite{Matthiesen2013} & InAs/GaAs QDs, \SI{951}{nm} & {---} & RF & EOM-carving, \SI{300}{MHz} \\
      \hline
    \textbf{This work} & InAs/GaAs QDs, \SI{933}{nm} & 515 MHz / 1 GHz & RF & EOM-carving, \SI{1}{GHz} \\
    \hline
  \end{tabular}
  \caption{Comparison of experimentally demonstrated high repetition rate excitation of quantum dot (QD) photon sources. DBR: distributed Bragg reflector, CBG: circular Bragg grating, SIL: solid immersion lens, FPC: Fabry-Perot cavity, PCW: photonic crystal waveguide, ABE: above-band excitation, RF: resonance fluorescence, TPE: two-photon excitation. }
  \label{tab1}
\end{table*}

\section{Photon flux}

\begin{figure}[htb!]
	\centering
	\includegraphics[width=.9\linewidth]{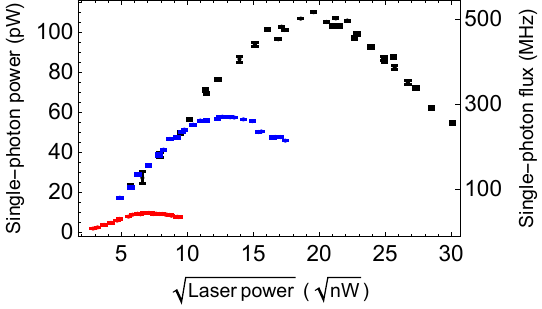}
    \caption{{\bf Powermeter-measured Rabi oscillations.} Single-photon optical power displaying Rabi oscillations as function of excitation laser power, at $80$~MHz (red), $500$~MHz (blue), and $1$~GHz (black) driving rate. A single-photon flux above $500$ MHz occurs at $\pi$-pulse drive at maximum excitation rate.}
 \label{fig3}
\end{figure}

A 100 \% efficient source of single photons at $933.1$~nm wavelength operated at a $1$~GHz photon rate generates an optical power of $212.9$~pW. We have recently reported (fibred) system efficiencies above $50\%$~\cite{Loredo2026}, which enables sufficiently high average optical powers that can be detected with off-the-shelf optical powermeters. Figure~\ref{fig3} displays intensity Rabi oscillations measured with a commercially-available powermeter at $80$~MHz, $500$~MHz, and $1$~GHz excitation rates. Their optical powers at $\pi$-pulse excitation are $9.57$~pW, $57.3$~pW, and $110$~pW, respectively, corresponding to directly measured efficiencies of $56.2\%$, $53.8\%$, and $51.5\%$ at the respective excitation rates. We attribute the slight decrease of efficiency at higher excitation rates to the increasing overlap between single-photon pulses, i.e. at higher driving rate the quantum dot does not fully decay before next excitation event. At $1$~GHz excitation, $\pi$-pulse drive hence produces $515$~MHz of in-fibre single-photon flux. For comparison, Table~\ref{tab1} lists other works of quantum dot based photon sources driven at high excitation rates.

\section{Conclusion}

We have combined GHz excitation rates and system source efficiencies above $50\%$ to produce more than $500$~MHz single-photons available in a single-mode fibre, the highest value of single-photon flux reported to date. The source produces an average optical power  that can be directly detected with standard photodiode powermeters, which serves as a direct way to quantify the fibred single-photon source efficiency. We note that by simply using a powermeter to measure the source fibre efficiency we can avoid the need for calibration of superconducting nanowire single-photon detectors, often involving a number of measurement-dependent assumptions and parameters, e.g. detector dead-times, bias currents, arriving flux, and photon-number dependent efficiencies. The source may find immediate applications as power standards in metrology or de-multiplexed for many-photon quantum simulation experiments comfortably in the $10-20$ photon range. We therefore anticipate a number of experimental photonic quantum information protocols being unlocked and enabled with this update of available solid-state single-photon sources.


\clearpage
\bibliography{ref.bib}
\end{document}